\documentstyle[11pt]{article}

\newcommand{\cA}{{\cal A}}
\newcommand{\cB}{{\cal B}}
\newcommand{\br}{{\bf r}}
\newcommand{\alf}{{(\alpha+1)(\alpha+2)}}
\newcommand{\be}{\begin{equation}}
\newcommand{\ee}{\end{equation}}
\newcommand{\ba}{\begin{array}}
\newcommand{\ea}{\end{array}}

\begin{document}

\begin{titlepage}

\pagestyle{plain}

\title{ 
{\bf Isospin dependence of nuclear matter symmetry energy coefficients
 } }
\author{  {\bf F\'abio L. Braghin} \thanks{e-mail: 
braghin@if.usp.br}  \\
{\normalsize 
 Nuclear Theory and Elementary Particle
 Phenomenology Group,}\\
{\normalsize 
Instituto de F\'\i sica, Universidade de 
S\~ao Paulo, C.P. 66.318; CEP 05315-970;  S\~ao Paulo - SP, Brazil. }\\
}

\date{}
\maketitle
\begin{abstract} 
Generalized 
symmetry energy coefficients
of  asymmetric nuclear matter are obtained as the 
screening functions.
The dependence of the isospin symmetry energy coefficient
on the neutron proton (n-p) asymmetry may be determined unless
by a constant (exponent) $Z$ which depend on microscopic properties.
The dependence of the generalized symmetry energy coefficients 
with Skyrme forces on the n-p asymmetry and 
on the density, only from .5 up to 1.5 $\rho_0$,
 are investigated in the isospin and scalar channels.
The use of  Skyrme-type effective forces 
allows us to obtain analytical expressions  for these parameters
as well as their dependences on the neutron-proton (n-p) asymmetry,
density and even temperature.
Whereas the density dependence of these coefficients obtained with 
 Skyrme forces
is not necessarily realistic the dependence on the 
n-p asymmetry exhibit a more consistent behaviour.
The isospin symmetry energy coefficient (s.e.c.) 
increases as the n-p asymmetry 
acquires higher values whereas the isoscalar s.e.c. decreases.
Some consequences for the Supernovae mechanism 
are discussed. 
\end{abstract}



IF- USP - 1476/2001

\noindent {\it To be published in Nucl. Phys {\bf A}}

\end{titlepage}

\section{Introduction}

The  symmetry energy coefficients are of
interest for the understanding  of 
several aspects of nuclear structure and also reactions as for instance
(neutron rich) nuclei structure,  giant resonances
 and  nuclear heavy ion collisions at intermediary and high 
energies \cite{TANIHATA,HAJENJON,PAK,FASRES,LIKOREN}.
There are also astrophysical motivations for this 
as, for example, in the supernovae mechanism and r-processes in 
nucleosynthesis \cite{BETHE,SUSUTO,PEARNAYA}.
However, these coefficients can depend on the medium properties
as density, temperature and neutron-proton asymmetry being these
dependences relevant for different phenomena.
Although one usually only considers the coefficient of the isospin
channel ($a_{\tau}$) 
there are other coefficients for the spin, spin-isospin and 
scalar channels of the effective nucleon nucleon (NN) interaction.
The former is usually  defined 
as the energy difference
between the unpolarized and polarized 
nuclear matter ($a_{\sigma}$) while the latter to the completely 
symmetric unpolarized  and
asymmetric polarized neutron matter ($a_{\sigma \tau}$).
However these two last parameters are believed to be small
and will not be considered in the present article.
The most well known and studied of these parameters is  the
isospin one ($a_{\tau}$) and we will refer to it unless 
explicitly quoting the others.
This parameter  measures the restoring force of the nuclear 
system to a perturbation which
 separates protons from neutrons.
It corresponds to a cost in energy which  however 
depends on the asymmetry of the medium.

In the macroscopic mass formulae 
the symmetry energy coefficient (s.e.c.) (in the isospin channel)  
 measures the difficulty of increasing the 
neutron-proton (n-p) asymmetry for a stable system.
It contributes as a  coefficient for the squared 
neutron-proton asymmetry: 
\be \label{1}
E/A = H_0(A) + a_{\tau}(N-Z)^2/A^2,
\ee
where $H_0$  does not depend on the asymmetry,
 Z, N and A are the proton, neutron and mass numbers respectively. 
An analogous form for the scalar channel will be considered yielding
a parameter related to the nuclear matter incompressibility.
Expression (\ref{1}) is obtained from the Fermi gas model
\cite{BETWEI} as well as from microscopic calculations \cite{BOMBALOM} 
but it is also known that there are important corrections
due to nucleon-nucleon interactions and correlations.
Higher orders effects of the asymmetry (proportional to $(N-Z)^n$
for $n \neq 2$,  n=1, 3) correspond to the explicit
breaking of isospin symmetry among other effects
\cite{JACO}. They are expected, in principle, 
to be less important 
for the equation of state (EOS) of nuclear
matter \cite{LKLB,MONI}.
However we show that there may occur relevant deviations 
for n-p asymmetries
by studying the generalized s.e.c. as 
the nuclear matter screening functions.

For the purpose of comparison we quote one work in which the 
density dependence of the isospin s.e.c. worked out
 \cite{PAL}. In this reference the relevance of $a_{\tau}$ for the 
equation of state (EOS) of dense neutron stars
was studied with an effective form for the 
nuclear interaction. 
In that work the following parametrization was obtained:
\be \label{11} \ba{ll}
\displaystyle{ a_{\tau}(\rho) = 
S(\rho) =  (2^{\frac{2}{3}}-1)\frac{3}{5}E_F^0
(u^{\frac{2}{3}} - F(u) ) + S_0 F(u) },
\ea
\ee
where $u=\rho/\rho_0$, $\rho_0$ being the saturation density of nuclear
matter, 
$E_F^0$ is the Fermi energy of nuclear matter,
$S_0$ is the $a_{\tau}$ at the saturation density (nearly 30 MeV)  
and $F(u)$ 
is a generic function which satisfies the condition $F(1) = 1$. 
With non relativistic microscopic  approaches,
 $a_{\tau}$ is found to vary nearly linearly
with the density until $\rho \simeq 2.5 \rho_0$ and 
to saturate at higher densities. 
However, this behaviour may depend on the nucleon-nucleon
potential \cite{BOMBALOM,EJEMMUP,WIRINGA}.

In the present article we perform a detailed investigation of 
 the nuclear matter s.e.c.
 as the screening functions of nuclear matter 
for the isospin and scalar channels.
Preliminary results were presented in \cite{FB2000a}. 
Now  we extend that calculation 
for the case of Skyrme forces with more 
involved density dependence \cite{DUTOABO,ONSIPP}
and make a quite complete analysis of them. 
Different  effective Skyrme forces are considered in order to 
assess the possible behaviour of these functions.
The use of these forces allows us to derive analytical
expressions for the generalized s.e.c.
One may argue that Skyrme effective interactions
are parameterized purely on phenomenological grounds and therefore would not
have predictive power for nuclear matter at densities different from
$\rho_0$ (the saturation density) or at very large n-p asymmetries.
In the present work, however, we consider different Skyrme forces
fitted from diverse sources including one derived from studies with 
realistic interactions \cite{CHABANAT}.
Therefore, a study with these different forces can supply some information
on the behaviour of such phenomenological models.
Furthermore, as Skyrme parameterizations are frequently used, this
kind of information can be relevant in diverse situations.
Besides that, a general calculation is done which yields a general 
expression for the isospin dependence of the
screening function.
The dependences of these generalized coefficients on the n-p asymmetry
and on the nuclear density (in a short range: from $0.5 \rho_0$ 
up to $1.5 \rho_0$) will be studied in this article.
The relevance of some results for the Supernovae (SN) mechanism
is discussed. 
In the next section the general expression for 
the screening functions is derived and 
the particular expression for
Skyrme forces is presented. The specific form of the functions
of the Skyrme parameters is shown in the appendix for each 
of the channels.
In section 3 the results are presented and discussed. 
In the final part the conclusions are summarized.

\section{Generalized Symmetry Energy Coefficients}

Generalized nuclear matter symmetry energy coefficients will be 
investigated in the following. 
It is interesting to review and to 
extend a qualitative argument  from  
\cite{BVA} for exploring them. 

\subsection{ General Remarks}

Let us consider a small amplitude ($\epsilon$) 
external perturbation which acts, through the third
Pauli isospin matrix  $\tau_3$, in nuclear matter 
 separating nucleons with isospin up and down 
\footnote{
This argument is valid for all
the four channels -isovector, spin, spin-isospin and scalar- with
suitable modification. It is enough to consider other external 
perturbations:  $\sigma$, $\sigma_3 \tau_3$ and {\bf 1} for 
the spin, spin-isospin and 
scalar channels respectively.}. 
This originates a fluctuation 
$\delta \rho = \rho_n - \rho_p$ of the nucleon density. 
The total energy of the system can be written as:
\be \label{1as}
\displaystyle{ H = H_0 + a_{\tau}\frac{\delta \rho^2}{\rho} +
\epsilon \delta \rho ,}
\ee
where $a_{\tau}$ is the isospin symmetry coefficient and $H_0$ does not
depend on the asymmetry.
In the equilibrium the following stability conditions must hold:
\be \label{stacon} \ba{ll}
\displaystyle{\frac{\delta H }{\delta \rho } =  2 \frac{a_{\tau}}{\rho} 
+ \epsilon = 0. }
\ea
\ee
The ratio of the amplitude of the fluctuation generated by the external
perturbation to the amplitude $\epsilon$ of this perturbation yields
the  static polarizability ($\Pi$)
(now generalizing for any channel 
as done in \cite{BRAVAU,FB99a} 
with (s,t) for spin,isospin numbers):
\be \label{12z}
\displaystyle{ 
\frac{\delta \rho_{s,t}}{\epsilon_{s,t}} \equiv \Pi^{s,t} 
= - \frac{\rho_0}{2 A_{s,t}}
,}
\ee
Where the $A_{s,t}$ are equal to the s.e.c.
In particular in the isovector channel: $A_{0,1} = a_{\tau}$.
This expression corresponds to the static limit of the response function
of symmetric nuclear matter \cite{BVA}.

Now let us consider asymmetric nuclear matter. In the usual 
parameterizations for the mass formulae, there are several
 effects which depend on the induced asymmetry 
$\delta \rho = \rho_n - \rho_p \equiv \beta$.
One can thus extend expression (\ref{1as}) to include a more general
dependence on the asymmetry. Our approach consists in 
considering a general parameter ${\cal A}$ which is also 
a function of the n-p asymmetry (and therefore a function of $\beta$)
instead of the
s.e.c. as usually defined in mass formulae 
of the type of (\ref{1as}). 
If the same external perturbation 
 is introduced in the energy density we can write:
\be \label{1asa}
\displaystyle{ H = H_0 + {\cal A}^{0,1}(\beta) \frac{\beta^2}{\rho} +
\epsilon \beta .}
\ee
It is worth to note that ${\cal A}^{s,t}$ (in a notation of a generic channel)
has rather a dependence on the n-p asymmetry $b = \rho_n/\rho_p -1$.
The exact form of this dependence is not the same that the 
dependence on the fluctuation $\beta$. 
In other words we could write, for the sake of completeness,
${\cal A} = {\cal A}(\beta, b)$.
However these two parameters should be related to each other.
In \cite{FB99a} two different prescriptions were discussed for
$\beta$ in the calculation of the response function of asymmetric 
nuclear matter. 
The most reasonable one leads to the following relationship 
between  the fluctuation $\beta$ and the asymmetry $b$:
\be \label{relac} \ba{ll}
\beta = \delta \rho_n \left( \frac{2+b}{1+b} \right),
\ea
\ee
Where $\delta \rho_n$ is the neutron density fluctuation. 
In the symmetric limit $\beta = 2 \rho_n$ and in the opposite limit, 
in neutron matter, $\beta = \delta \rho_n$, as expected.
Considering this, we are lead to expect that the dependence of 
${\cal A}$ on the asymmetry $b$ is related to its dependence on the
density fluctuation $\beta$. This allows us to write
shortly: ${\cal A} = {\cal A}(\beta)$. 
It is important to stress that the prescription (\ref{relac})
is model-dependent and different choices for it yield different 
results for the static screening functions as shown below.
The  dynamic response functions are less sensitive to this
prescription, but not completely independent \cite{FB99a}.
The above prescription (\ref{relac}) is based on the assumption that
the density fluctuations are proportional to the respective 
density of neutrons and protons, i.e., 
$\delta \rho_n/ \beta = \rho_n / \rho$, being $\rho$ the total 
density.
In spite of being rather well suited for the isovector channel, 
this kind of assumption can be considered as a simplified model for the 
other spin/isospin channels in asymmetric nuclear matter.
The derivative of ${\cal A}$ 
with relation to $\beta$ is therefore  related to
$ d {\cal A}/d b$. 

If the system is stable with relation to the 
density fluctuations induced by the external source,
the following equation holds:
\be \label{deriv} \ba{ll}
\displaystyle{ \frac{\delta H}{\delta \beta} = 2 \cA \frac{\beta}{\rho} + 
\frac{\delta \cA}{\delta \beta} \frac{\beta^2}{\rho} + \epsilon =0}.
\ea
\ee
By substituting the variables  $\Pi \equiv \beta / \epsilon$ and 
$\cB = \cA / \rho$ in this equation  we re-write:
\be \ba{ll}
\displaystyle{ \frac{\delta \cB }{\delta \Pi} = -\frac{1}{\Pi^2} -
\frac{2 \cB}{\Pi}. }
\ea
\ee
We can face this expression as a first order differential equation
whose most general solution is given by:
\be \label{solut} \ba{ll}
\displaystyle{ \cA = - \frac{ \rho( \Pi - C)}{\Pi^2},
}
\ea
\ee 
where $C$ is a constant. 
From this expression we find the following algebraic equation:
\be \ba{ll}
\displaystyle{  \Pi^2 \frac{\cA}{\rho} + \Pi - C = 0
,}
\ea
\ee
where $C$ is a constant. The solutions are:
\be \label{solex} \ba{ll}
\displaystyle{ \Pi = \rho \frac{ -1 \pm \sqrt{1+ 4 C \cA/\rho}}
{2 \cA}. }
\ea
\ee
The constant $C$ is chosen in such a way 
that, in the limit of symmetric nuclear matter, the polarizability
$\Pi $ is the one calculated in expression (\ref{12z}). This
leads to $C= -\rho/(4 \cA_{sym})$. 

A safer and sounder way of obtaining solutions for the polarizability
is to find
the value of $\beta$ at which expression (\ref{deriv}) is satisfied.
We obtain a quadratic equation for $\beta$
whose solutions are given by:
\be \label{1at} \ba{ll}
\displaystyle{ 2 \beta \frac{\delta \cA^{0,1}}{\delta \beta} =
-2 \cA^{0,1} \pm 2\cA^{0,1} \sqrt{ 1 - 
\frac{\epsilon \rho}{{\cA^{(0,1)}}^2}
\frac{\delta \cA^{0,1}}{\delta \beta}  } .  }
\ea
\ee
There are two solutions for this expression and we will be concerned 
only with one of them (which produces sounder results) in the limit of very 
small amplitude $\epsilon$. 
In this limit for the positive sign of expression (\ref{1at}) 
the solution is trivially found to be:
\be \label{1bs}
\Pi^{s=0,t=1} \equiv \frac{\beta}{\epsilon} = 
- \frac{\rho}{2 \cA^{0,1}}.
\ee
This expression has exactly the same form of  symmetric nuclear matter.
The difference is that, here, $\cA$ does depend on $b$.

It is also possible to extract an explicit isospin dependence of the 
generalized symmetry energy coefficient.
From the solution of the polarizability (\ref{1bs}) we consider
the first derivative with relation to $b$:
\be \label{DERIV1} \ba{ll}
\displaystyle{ \frac{d \beta}{d b} = \frac{\epsilon \rho}{2 \cA^2 } 
\frac{d \cA }{d b} = - \frac{\beta}{\cA} \frac{d \cA}{d b}. }
\ea
\ee
A complementary expression can be obtained from the relation between
$b$ and $\beta$ of (\ref{relac}). It yields:
\be \label{relder} \ba{ll}
\displaystyle{ \frac{d \beta}{d b} = - \frac{\beta}{2+ b}.}
\ea
\ee
Equating these two last equations we obtain:
\be \label{relder2} \ba{ll}
\displaystyle{  - \cA \frac{\beta}{2+ b} = - \beta \frac{d \cA}{d b}
,}
\ea
\ee
From which it is possible to  derive the following relation 
between the isospin s.e.c. and the n-p asymmetry:
\be \label{isosb} \ba{ll}
\displaystyle{ \cA = \cA_{sym} \left(\frac{(2 + b)}{2} \right)^Z ,}
\ea
\ee
where $Z$ is a constant which depends on the channel (isovector in the 
present case) and eventually on other properties of the system
 and $\cA_{sym}$ is the 
s.e.c. of symmetric nuclear matter.
In the symmetric case $b=0$ and hence $\cA=\cA_{sym}$. 
For $b= 2$ (neutron density three times larger than the
proton density) we obtain $\cA = 2^Z \cA_{sym}$.
If we consider $Z=1$ a considerably high value is obtained. 
It is therefore missing some dynamical information to 
fix the constant $Z$ for each channel. 
In the next section we find that different Skyrme forces
will result different values of $Z$.
We want to emphasize that the prescription 
(\ref{relac}) was the relevant information for this calculation. 
Any other relation between $b$ and $\beta$
will induce different asymmetry dependence of the 
symmetry energy coefficients. 

Besides that  expression (\ref{1bs}) corresponds
to the static limit of the dynamical polarizability \cite{FB2000a}.
It defines the static screening functions $\cA^{s,t}$ in the usual way:
$$\epsilon \equiv C' \cA^{s,t} \beta,$$
where $C' =1/\rho_0$.
From this expression we see that
the external field $\epsilon$ induces density fluctuations
$\beta$ according to the intrinsic properties of the medium, given by 
the screening function $C' \cA^{s,t}$.
These arguments are valid for any 
 perturbation of the other channels for asymmetric nuclear
matter in isospin as well as in scalar channels yielding the functions
${\cal A}^{s,t}$.

We have shown that the polarizabilities of asymmetric
nuclear matter yield the generalized symmetry energy coefficients.
The dependence of the s.e.c. on the n-p
asymmetry can be deduced unless for a constant exponent $Z$. 
It is however necessary to provide a prescription for the
induced fluctuation $\beta$.
A nearly exact expression for the dynamical polarizability 
of a non relativistic
hot  asymmetric nuclear matter for variable 
densities was derived with Skyrme interactions in \cite{FB99a}. 
Its static limit is exploited in the following sections
for more general Skyrme forces.

\subsection{ Screening functions with Skyrme forces }

In \cite{FB99a,FB2000a} the dynamical polarizabilities of 
asymmetric nuclear matter were calculated using Skyrme forces.
An alternative and independent calculation for the dynamical polarizability
 was done in \cite{NAVARRO}.
Although the static limit has been found yielding the
screening functions it has not been 
exploited sufficiently well. 
It can also be derived directly
from the static Hartree-Fock equation with the same external 
perturbation. 
The static generalized symmetry energy coefficients of 
hot asymmetric nuclear matter, 
$\cA_{s,t}$, can be written as:
\be \label{20} \ba{ll}
\displaystyle{\cA_{s,t} \equiv - \frac{\rho}{2 \Pi_{s,t} }
\frac{\rho}{ N}
\left\{ 1 + 2\overline{V_0^{s,t}} N_c 
+6 V_1^{s,t}M_p^* ( {\rho}_c + {\rho}_d ) + 
12M^*_p V_1^{s,t} \overline{V_0^{s,t}} \left( N_c {\rho}_d
 - {\rho}_c {N}_d \right) + \right.  } \\
\displaystyle{ \left. (V_1^{s,t})^2 \left( 36(M^*_p)^2 {\rho}_c
{\rho}_d - 16M_p^* M_c  N_d \right)  \right\} }.
\ea
\ee
This expression was derived in \cite{FB2000a} and 
is one of the main concerns of the present article \cite{MORAWETZ}. 
The densities $\rho_v$, $N_v$ and $M_v$ are given by:
 $$ \rho_v = v \rho_n + (1-v) \rho_p,$$ 
$$ M_v = v M_n + (1-v) M_p,$$ 
$$ N_v = v N_n + (1-v) N_p,$$ 
where $v$ stands for n-p asymmetry coefficients ($c,d$)
defined below (a measure of the fraction of neutron density). 
The above densities are defined by:
$$ (N_q,\rho_q,M_q) = \frac{2M^*_p}{\pi^2}
\int d f_q(k) (k, k^3, k^5).$$
In these expressions $d f_q(k)$ are the differential 
fermion occupation numbers for neutrons ($q=n$) and 
protons ($q=p$). In the zero temperature limit they reduce
to delta functions at the Fermi surface momentum
$d f_q \to - \delta (k-k_F)$. At non zero temperature one has
to fix the chemical potential to calculate the Fermi occupation
number and the densities. 
But this study of the s.e.c. dependence on the temperature
would be more appropriately done
with an equation of state and it is beyond the scope of this article.
 $\overline{V_0}$ and $V_1$ are functions of the
Skyrme forces parameters (calculated for Skyrme forces with 
more general density dependence and shown in the appendix) and
$M^*_p = m^*_p/(1+a/2)$ is an 
effective mass for the proton. Besides that, the
four asymmetry coefficients are:
\be \label{5b} \ba{ll}
\displaystyle{
a = \frac{m^*_{p}}{m^*_{n}} - 1 , \;\;\;\;\;
b = \frac{\rho_{0n}}{\rho_{0p}}  -1, \;\;\;\;\; 
c = \frac{1+b }{2+b }, }\;\;\;\;\;
\displaystyle{ d = \frac{1 }{1+ (1+b)^{\frac{2}{3} } } }.
\ea
\ee
(The coefficient $b$ is related to a frequently used
asymmetry coefficient:
$\alpha = (2\rho_{0n}-\rho_0)/\rho_0$, 
by the expression: $b = 2\alpha / (1- \alpha ) $).

In the scalar channel the 
screening function will be referred to as the dipolar 
incompressibility, $K_D$, and is related to the usual nuclear
matter incompressibility, $K_{\infty}$ by the following expression
in terms of the SLyb Skyrme force  parameters \cite{CHABANAT}:
\be
\displaystyle{ K_D = K_{\infty} + \frac{4}{5}T_F - 2V_1k_F^2\rho_0
+ \frac{3}{4}t_3 \rho^{\alpha+1}_0 .}
\ee
This relationship is slightly different for the SkSC interactions
\cite{ONSIPP}.
In this reference the authors 
fitted neutron star properties with 
parameterizations that yielded different values 
for the n-p symmetry energy at saturation 
density: SkSC4 ($27 MeV$), SkSC6 ($30 MeV$) and SkSC10 ($32 MeV$).
At different densities and n-p asymmetries (and also temperatures) 
the static
screening functions have different values. They are studied in the
next section.

\section{Results and discussion} 

In this section we show some figures which exhibit the 
isospin and density dependences of the nuclear matter screening functions
of expression (\ref{20}) for the isospin and scalar channels. 
For this we use the functions $\overline{V_0}$ and $V_1$
(written in the appendix)
calculated for Skyrme functions with different density dependences.
The parameters of one of them (SLyb) were fitted from  results 
of neutron matter properties obtained from microscopic calculations 
in \cite{CHABANAT}. 
Other forces (SkSC4, SkSC6 and SkSC10), which
have slightly different density dependences, had their parameters
fixed by adjusting a large amount of nuclear masses yielding
the same results of the Extended Thomas Fermi method and
shell corrections calculated by the Strutinsky-integral method
\cite{DUTOABO,ONSIPP}. 
One of the strong characteristic of the forces SkSC
 is the value of the 
effective mass which is kept to be equal to the
free nucleon mass. 
This may be interpreted as a result of the coupling 
between particle modes and surface vibration modes. 

In figure 1 the generalized isospin symmetry energy 
coefficient $\cA_{0,1}$ is shown
as a function of the ratio of the density to the saturation density 
for Skyrme interactions 
SLyb,  SkSC4, SkSC6 and SkSC10. 
Since Skyrme forces are not necessarily expected 
to describe physics at high densities we decided to
investigate the dependence of the screening functions up to 1.5 $\rho_0$.
Most of the isovector screening functions of figure 1
does not vary much. Most of them have the tendency to keep nearly 
constant in a 
value close to that at the saturation density and decrease at higher
densities. For the SLyb force two curves are shown. 
$\cA_{0,1}$ corresponds to the complete s.e.c. expression (\ref{20}) 
whereas $a_{\tau}$ neglects the $V_1$ term present in expression 
(\ref{append1}). They exhibit nearly the same behaviour.
The SkSC10 force, however, yields a different behaviour which
is more compatible with that of the parameterizations of 
expression (\ref{11}),  with $F(u)=u$ from \cite{PAL} (plotted with circles). 
There are no  means (not yet) of choosing the most  realistic (and reliable)
of the forces in this channel for varying densities.
The force SkSC10 reproduces rather the behaviour of 
relativistic parameterizations.
In figure 2 the generalized isospin  
s.e.c. $\cA_{0,1}$ is shown as a function
of  the n-p asymmetry $b$ for the different forces at the saturation
density $\rho_0$.
There is a  trend which is absolutely dominant for all the interactions.
 It is the increase of $\cA_{0,1}$ with $b$.
One big difference between these curves for each of the forces  is given by 
the slope which can be larger (eg. for SkSC6) or smaller (for 
SLyb). It seems to us that if the slope were so large as that obtained from
SkSC6 it would have already manifested in studies of neutron matter
as well as in heavy nuclei observables (for the lead nucleus $b=0.54$).
Further studies of neutron stars, supernovae and heavy ions collisions 
may  lead to a more  precise determination of this slope.
In section 2.1 a general 
expression (\ref{isosb}) for the isospin dependence of the isospin s.e.c. 
has been derived making use of relation (\ref{relac}).
However it was not possible to fit any of the results obtained 
using the above Skyrme forces with expression (\ref{20}), for any $Z$.
The form of the curves (slopes) obtained with the used Skyrme forces, 
in figure 1, indicates that
$Z$ would be smaller than one whereas the numerical values 
of $\cA_{0,1}$ (with the same Skyrme interactions) suggest that $Z \geq 1$.
This indicates that the calculation done in section 2 is not completely 
compatible with the Skyrme forces which have been used in this paper.
This could eventually be solved by means of microscopical calculations 
(e.g. using Monte Carlo techniques as done in \cite{DEAN}) and could,
hopefully, also be checked by new parameterizations of mass formulae.

Another issue which can be taken from figures 1 and 2 is the 
isospin dependence of the s.e.c. at diverse densities.
The effects of changing the asymmetry and the density sum up 
(with relation to the symmetric nuclear matter at saturation density)
reproducing behaviours already present in figures 1 and 2 for given
$\rho$ and $b$. 
This means that the same results of figure 1 (density dependence)
will be present for  nuclear matter with an asymmetry b (shown in 2). 
The difference is that: the more asymmetric is the nuclear matter
more repulsive is the isovector interaction (and hence 
higher the $\cA_{0,1}$).

In \cite{ONSIPP} the EOS of a collapsing star 
is studied and the results
support forces for which $a_{\tau}$ is considerably greater
than $27 MeV$ to keep neutron matter stable. 
However these higher values spoil the fit of nuclear masses
within the frame of the extended Thomas-Fermi plus Strutinsky integral mass
formula \cite{PEARNAYA}. 
Our results make these discrepancies be compatible in a certain sense.
We have found that for larger values of the n-p asymmetry $b$ 
the isospin s.e.c. increases considerably
corresponding to the case of neutron stars medium.
Therefore for the same Skyrme force several values of $a_{\tau}$, or 
rather $\cA_{0,1}=\cA_{0,1}(b,\rho)$, can be considered. 
For the description of 
(not very asymmetric) nuclei the lower values can be expected to yield
better results.

For the generalized scalar dipolar s.e.c., 
figures 3 and 4 show that, roughly speaking, all the interactions
are attractive for lower densities (figure 3) and higher n-p asymmetries
(figure 4). 
This corresponds to instabilities which are related to 
those found, for example, in \cite{AYCOCHO}.
For neutron stars it may seem that all these forces (except SLyb) lead
to instabilities according to figure 4 (for fixed density $\rho_0$). 
However higher densities
may stabilize the system (while lower densities lead to still more unstable
systems).
As discussed for the isospin case the effects of varying the 
nuclear density and the asymmetry can be considered simultaneously,
i.e., to obtain the n-p asymmetry dependence of the scalar 
s.e.c. at a density $\rho \neq \rho_0$ the behaviours of figures
3 and 4 can be analysed together.
Therefore, for higher densities (making the $\cA_{0,0}$ to increase)
the system may experience higher asymmetries ($b$) without the
instabilities present for forces SkSC4 and SkSC6. 
We do not exhibit, 
however, a quantitative estimate of the instabilities points in the variables
 $b$ and $\rho$ due to
the diverse behaviour of Skyrme forces. Moreover these points may be
out of the validity of the application of these interactions. 
For the moment there is no way of verifying this.
The interaction in this channel becomes more and more repulsive 
for higher densities. 
The increase of the n-p asymmetry on the other hand,
make the interaction continuously more attractive 
for most of the forces as shown in figure 4.
However, the slopes for interactions SkSC seem to be too negative.
In  figure 4 it is also shown  
the parameterization for n-p asymmetry dependence of 
the incompressibility of nuclear matter ($K_{\infty}(b)$)
of reference \cite{BCK} -from the equation of state- 
(not the incompressibility $K_{\infty}$ itself, 
but its asymmetry dependence- for the case of SLyb).
This parameterization (using the n-p asymmetry parameter of our work) 
is given by: 
\be \label{18a}
K(b) = K(b=0) (1- a' (b/(b+2))^2 )
\ee
where $a'$ (in \cite{BCK}) is of the order of 1.28 up to 1.99 
for several Skyrme interactions (zero temperature)
and $K(b)$ stands for $K_{\infty}$ or $\cA_{0,0}=K_D$. 
It is seen in figure
4 that this case (from \cite{BCK}) has
nearly the same behaviour (dependence on the n-p asymmetry b
with $\rho=\rho_0$) as our calculation for force SLyb,
a remarkable feature.
In another work \cite{ONSIPP} it was found the same kind of
dependence on the
n-p asymmetry for the incompressibility of nuclear matter but with other
numerical coefficients:
\be
K(b) = K_v + K_s (b/(b+2))^2 
\ee
These parameters assume the following values for the forces used in that
article:
\be \ba{ll}
 SkSC4 (K_v=234.7 MeV, \;\; K_s=-334.9 MeV ); \;\;\;\;\;\;
   SkSC10 (K_v=235.4 MeV, \;\; K_s=-203.5 MeV ); \\
  SkSC6 (K_v=235.8 MeV, \;\; K_s=-136.6 MeV ). 
\ea
\ee
These values are (a little smaller) of the same order   
of those obtained in the figure 4 and expression (\ref{18a}).
It can  expected that this dipolar 
incompressibility is also  of relevance for 
the study of the Isoscalar Dipole Giant Resonance \cite{ISDGR}.

Consequences concerning the supernovae mechanism can be discussed
now.
The higher the n-p symmetry energy coefficient the smaller is the
deleptonization (electron capture) 
in the quasi-static phase of the supernovae mechanism 
yielding a larger final proton fraction 
and faster cooling (via neutrino emission). 
As we have found that $\cA_{0,1}$ increases with the n-p asymmetry
it can be expected that as deleptonization proceeds it becomes more
and more difficult to convert protons into neutrons.
This also means that the final 
(neutronized) star would be less asymmetric than
expected with smaller (fixed) values for $a_{\tau}$.
This picture yields a stronger shock wave since the energy 
loss inside the star due to deleptonization is smaller.
Therefore the increase of isospin symmetry energy
coefficient (which occurs for higher n-p asymmetries)
helps a successful explosion of the (contracting) star (keeping the density
fixed).
This conclusion has the essential feature  compatible
with the analysis done for the SN 1987a event 
\cite{BETHE,DEAN,DONATI}.

\section{ Conclusions}

Summarizing, generalized symmetry energy coefficients
of nuclear matter 
were investigated.
Firstly 
we have shown that the polarizabilities of asymmetric
nuclear matter yield the generalized symmetry energy coefficients.
The dependence of the isospin s.e.c. on the n-p
asymmetry, by means of the relation (\ref{relac})  
was found to be: 
$$\cA = \cA_{sym} \left(\frac{(2 + b)}{2} \right)^Z ,$$
where $Z$ is a constant. 
To obtain such expression it was necessary to provide a prescription for the
dependence of the induced fluctuation $\beta$ on the n-p asymmetry $b$ 
(expression (7))
\cite{FB99a} which could be chosen differently.
The form of the curves obtained from the used Skyrme forces in figure 1
indicate that
$Z$ would be smaller than one whereas the numerical values 
of $\cA_{0,1}$ suggest that $Z \geq 1$.
This constant seems to contain
relevant dynamical information and it could be determined
by the study of the symmetry energy coefficient microscopically.

Secondly a study of the generalized 
symmetry energy coefficients (isovector and scalar)
was done with Skyrme forces.
However these results are not adjusted by the function obtained
(the expression (\ref{isosb})) 
The use of Skyrme-type interactions allowed 
to obtain analytical expressions for the s.e.c. 
Their density and n-p asymmetry dependences were analyzed for 
different Skyrme forces which may yield very different 
behaviours. 
In some cases it is possible to discard unphysical 
results but there are results which cannot be chosen 
due to non existing experimental knowledge. 
Nevertheless neutron stars and eventually very asymmetric nuclei
properties as well as heavy ion collisions
can provide valuable information.
Although one should not believe that only one  
Skyrme force parameterization could account for the description
of all nuclear observables at different 
n-p asymmetries (as well as densities and temperatures) 
it is acceptable the idea that several parameterizations
could hopefully describe different ranges of the dependence of 
nuclear observables (as the s.e.c.)
with these three variables.
Nevertheless it is tempting to conclude that the force SLyb,
 which was adjusted to reproduce general properties
of asymmetric nuclear matter, yielded, in general,
more reasonable behaviour for the symmetry energy coefficients
for varying n-p asymmetries.
The dependence of the isospin s.e.c.
on the n-p asymmetry indicates that  the deleptonization process
in supernovae is suppressed as the neutronization occurs. 
This would make final proton fraction bigger with less
energy loss helping the SN explosion (and also less asymmetric neutron 
stars matter) in agreement
with expectations. A quantitative estimate of this effect however will 
only be possible after a more precise determination of the 
dependence of the s.e.c. on $b$.

\vskip 0.3cm
\noindent {\Large {\bf Acknowledgements}}

This work was supported by FAPESP, Brazil. 
\vskip 0.2cm

\vskip 0.5 cm
\setcounter{equation}{0}
\renewcommand{\theequation}{A.\arabic{equation}}
\noindent {\Large {\bf Appendix: Functions $V_i^{s,t}$ }}
\vskip 0.2cm

We write the parameters $V_0$ and $V_1$ from the 
linearization of the Hartree Fock equation (for nuclear matter) for the 
Skyrme forces SkSC used by \cite{ONSIPP}. They contain more
involved density dependence:
\be \label{append1} 
 \overline{V_0}^{s,t} = V_0^{s,t} + V_2^{s,t} - 
\frac{2 V_1^{s,t}}{1-4V_1^{s,t}m^*_p \rho_{0,p} },
\ee
where  only the $V_0$ and $V_2$ were modified.

In the SkSC forces the density dependence implies that this part of the 
interaction between two protons depends only on the proton density
being more reasonable.
It can be written in the following form:
\be \ba{ll}
\displaystyle{\frac{t_3}{6} (1 + x_3 P_{\sigma})
\left( a_1(\rho_{qi} +\rho_{qj})^{\alpha} \delta(\br_{i,j} )
+ a_2 \rho^{\alpha} \right),
} 
\ea
\ee
where $P_{\sigma}$ is the two-body spin exchange operator, $q$ indicates
proton or neutron and
$a_1, a_2$ are taken to be $0, 1$ or $1, 0$ for Skyrme interactions
SkSC and SLyb respectively.

In the isovector channel one obtains:
\begin{equation} \label{7}\begin{array}{ll}
&\displaystyle{
V_0^{0,1}= \left(
- \frac{t_0}{2} \left(x_0+ \frac{1}{2} \right) - \frac{t_3}{12}
\left[ a_2 \left( x_3 + \frac{1}{2} \right) 
+ a_1 \left(1 + \frac{x_3}{2}\right) -\frac{1}{4}(1-x_3)(\alpha+2)
(\alpha+1) \right]
 \rho_0^{\alpha} 
 \right) (1 + bc)},\\
&\displaystyle{ V_1^{0,1}
= \frac{1}{16} ( t_2( 1+ 2 x_2) -  t_1( 1 + 2 x_1) )}, \\
&\displaystyle{ V_2^{0,1} = t_3\left[ 
a_2 (\frac{1}{2}+x_3)\alpha \rho_0^{\alpha-1} (c\rho_n + (c-1)\rho_p ) +
\right. } \\
& \displaystyle{ \left.
+ a_1 \left( (1+\frac{x_3}{2})\alpha \rho^{\alpha-1} (c\rho_n + \rho_p(c-1) )
 + 2 (1- x_3) (\alpha+2) (\alpha+1) (c\rho_n^{\alpha} + 
\rho_p^{\alpha} (c-1) )/16
\right)
 \right] /12   ,}
\end{array}
\end{equation}
where $\rho_{0n}$,  $\rho_{0p}$  and $ \rho_0$ are the 
proton, neutron  and total saturation 
densities of  asymmetric nuclear
matter. For the scalar  channel:
\be \ba{ll}
 \displaystyle{
V_0^{0,0}= \left( 3 \frac{t_0}{4} + (\alpha+1)(\alpha+2) t_3
\rho_0^{\alpha} \left[ a_1 (1+\frac{x_3}{2})
\left(\frac{1+ b}{2+b}\right)^2\frac{1}{16} + a_2 (1+ \frac{x_3}{2})
\frac{1}{12} \right]
\right)(1+bc)
},\\
 \displaystyle{ V_1^{0,0}
=  3\frac{t_1}{16} + (5 +4 x_2) \frac{t_2}{16}
}, \\
 \displaystyle{ V_2^{0,0} = \frac{t_3}{12} \left[ 
(x_3+.5)(c \rho_n + (c-1)\rho_p
 \rho_0^{(\alpha-1.)}) +  \right. }\\
\displaystyle{ \left. +
a_1 \alf (1-x_3)\left( 
\frac{(2\rho)^{\alpha}}{(2+b)^{\alpha+2}} + 
2\frac{((1+b)^2\rho)^{\alpha}}{(2+b)^{\alpha+2}}  \right)/2
-a_2 ( 1 + (1+b)^2\rho^{\alpha} )/(2+b)^2 \right]
.}
\ea
\ee

\newpage

\noindent {\Large {\bf  Figure caption}}

\vskip 0.5cm

{\bf Figure 1} Isovector screening function 
 $A_{0,1}= \rho/(2 \Pi_R^{0,1})$ of symmetric nuclear matter
 as a function of the ratio
of density to density at saturation ($u=\rho/\rho_0$) for interactions SLyb
(solid),
SkSC4 (dotted), SkSC6 (dashed), SkSC10 (long dashed). 
The simplified expression ($a_{\tau}$), i.e. without
terms of order of $V_1^2$,  for the
force SLyb (thin dotted-dashed).
Circles (P.A.L. 1) for the expression 
of $a_{\tau}$ as a function of $u$
 with three different functions $F(u)$ from reference \cite{PAL}.

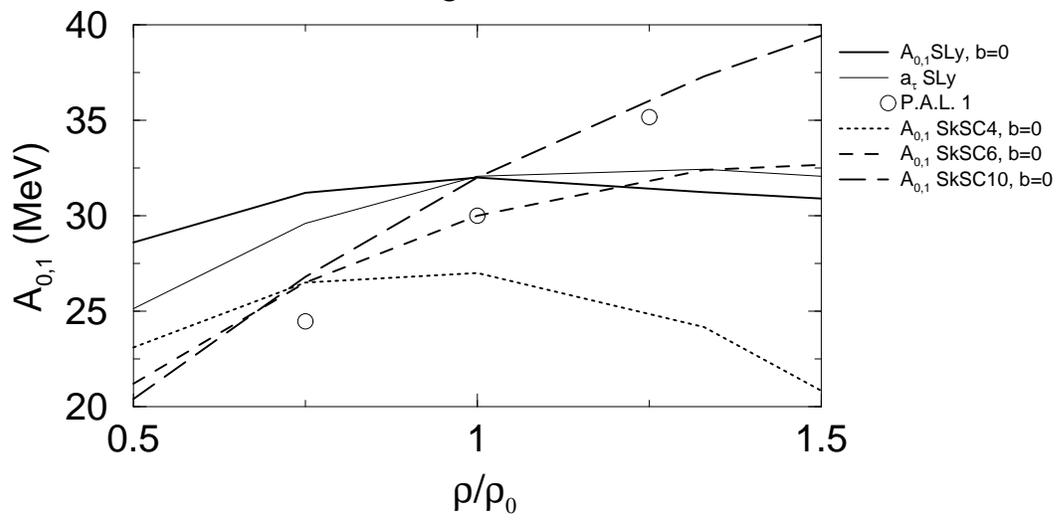
{\bf Figure 2}  Isovector screening function
$A_{0,1}= \rho/(2 \Pi_R^{0,1})$ as a function of the
asymmetry coefficient $b$ (at $\rho_0$): 
dotted line for SkSC4, dashed lines for SkSC6, solid line for SLyb
and long-dashed lines for SkSC10.

{\bf Figure 3} Scalar (dipole) screening function
 $A_{0,0}= \rho/(2 \Pi_R^{0,0})$
as a function of the ratio
of density to density at saturation ($u=\rho/\rho_0$) for the
different interactions with the conventions of figure 1.

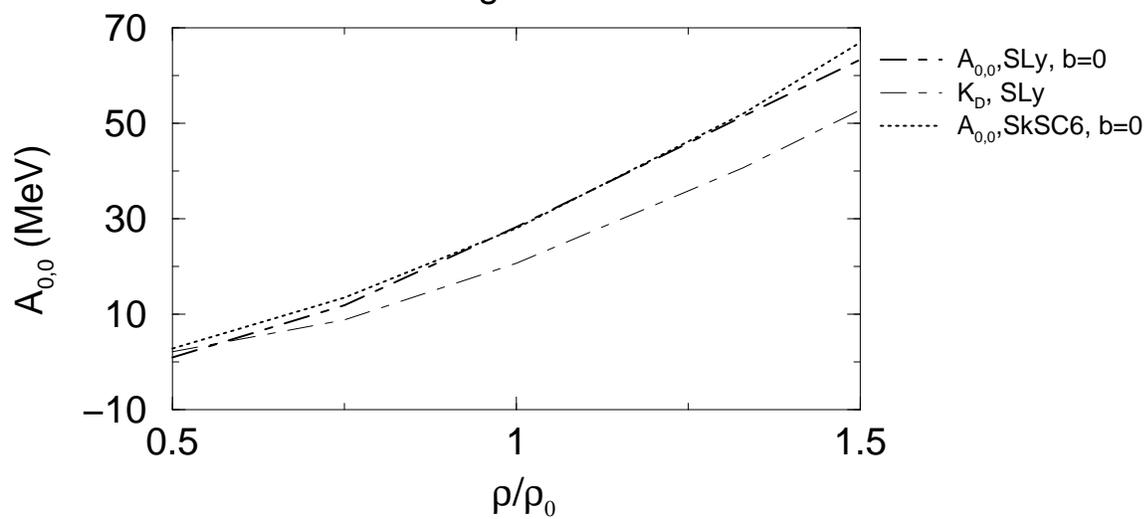
{\bf Figure 4}  Scalar (dipole) screening function
 $A_{0,0}= \rho/(2 \Pi_R^{0,0})$ as a function of the asymmetry
coefficient $b$, with $\rho_0$,  with the different forces
(the conventions of figure 2) and also the n-p asymmetry 
dependence of the compressibility $K_{\infty}$ of nuclear matter
of reference \cite{BCK} using $K_D(b=0)$ from the SLy force (star).


\begin{thebibliography}{ll}
\bibitem{MORAWETZ} Extensions of the nuclear matter dynamical polarizabilities
                  were calculated to include other
                  effects. See for example:
                    K. Morawetz, U. Fuhrmann, nucl-th/9907011.
\end{thebibliography}
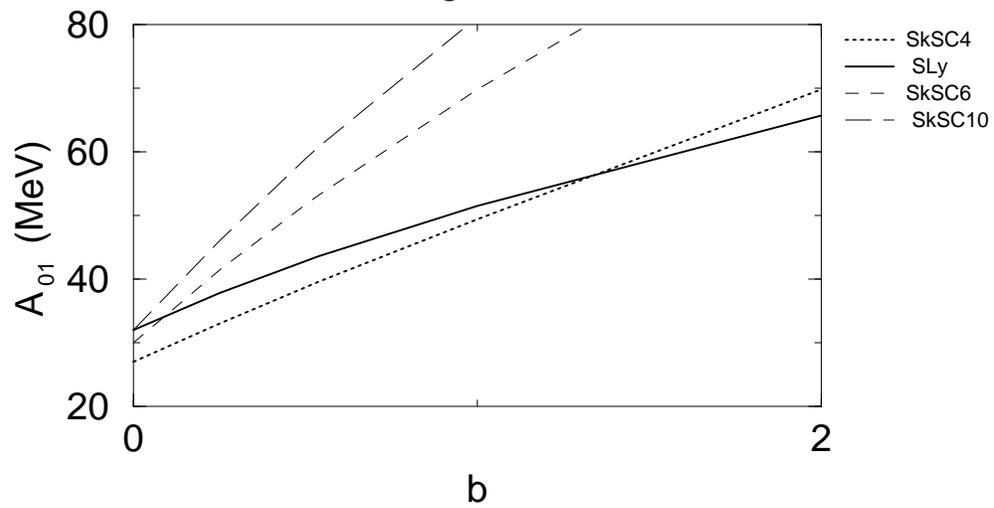
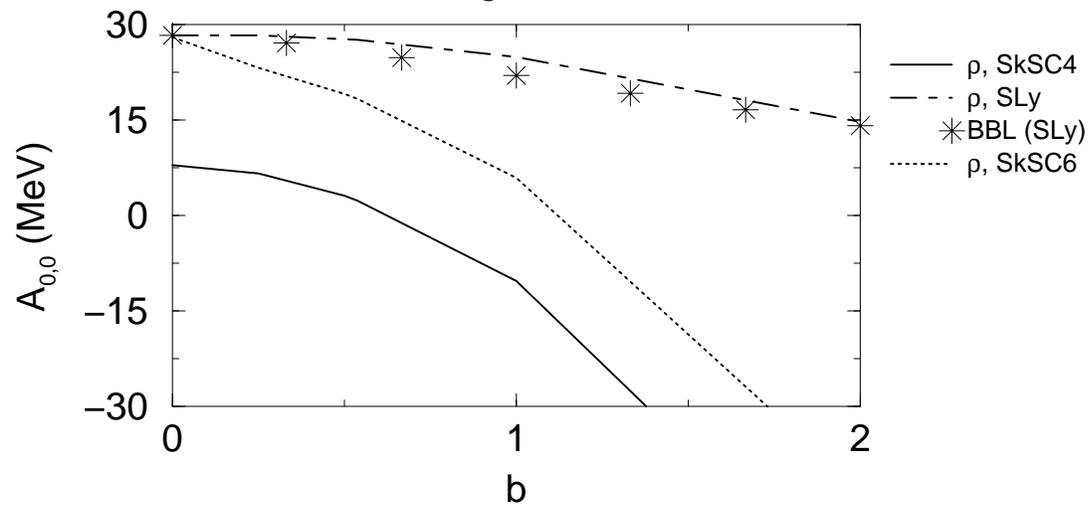
\end{document}